\documentclass[superscriptaddress,aps,prl,twocolumn]{revtex4}
\usepackage{bm}
\usepackage{ulem}
\usepackage{epsfig}
\usepackage{graphicx}
\usepackage{amssymb,amsmath,amsbsy,amsgen,amsfonts}    
\usepackage{dcolumn}
\usepackage{amsthm}
\usepackage{mathrsfs}
\usepackage{latexsym}
\usepackage{array}
\usepackage{color}
\usepackage{amstext}
\allowdisplaybreaks[1]
\usepackage{txfonts}
\usepackage{pifont}

\usepackage{epstopdf} %for texshop on mac

\newcommand{\bra}[1]{\left\langle{#1}\right\vert}
\newcommand{\ket}[1]{\left\vert{#1}\right\rangle}

\newcommand{\be}{\begin{equation}}
\newcommand{\ee}{\end{equation}}
\newcommand{\ba}{\begin{array}}
\newcommand{\ea}{\end{array}}
\newcommand{\bqa}{\begin{eqnarray}}
\newcommand{\eqa}{\end{eqnarray}}

\usepackage{color}
\definecolor{wm}{rgb}{0.3,0.0,0.9}

\DeclareSymbolFont{symbols}{OMS}{cmsy}{m}{n}

\begin{document}

\title{Experimental demonstration of a measurement-based realisation of a quantum channel}

\author{W. McCutcheon}
\affiliation{Quantum Engineering Technology Laboratory, Department of Electrical and Electronic Engineering, University of Bristol, Woodland Road, Bristol, BS8 1UB, UK}
\author{A. McMillan}
\affiliation{Quantum Engineering Technology Laboratory, Department of Electrical and Electronic Engineering, University of Bristol, Woodland Road, Bristol, BS8 1UB, UK}
\author{J. G. Rarity}
\affiliation{Quantum Engineering Technology Laboratory, Department of Electrical and Electronic Engineering, University of Bristol, Woodland Road, Bristol, BS8 1UB, UK}
\author{M. S. Tame}
\email{markstame@gmail.com}
\affiliation{School of Chemistry and Physics, University of KwaZulu-Natal, Durban 4001, South Africa}

\date{\today}

\begin{abstract}
We introduce and experimentally demonstrate a method for realising a quantum channel using the measurement-based model. Using a photonic setup and modifying the bases of single-qubit measurements on a four-qubit entangled cluster state, representative channels are realised for the case of a single qubit in the form of amplitude and phase damping channels. The experimental results match the theoretical model well, demonstrating the successful performance of the channels. We also show how other types of quantum channels can be realised using our approach. This work highlights the potential of the measurement-based model for realising quantum channels which may serve as building blocks for simulations of realistic open quantum systems.
\end{abstract}

%\pacs{}

\maketitle

%%%%%%%%%%%%%%%%%%%%%%%%%%%%
%%%%%%%%%%%%%%%%%%%%%%%%%%%%
%%%%%%%%%%%%%%%%%%%%%%%%%%%%
%%%%%%%%%%%%%%%%%%%%%%%%%%%%

{\it Introduction.---} The modelling and simulation of quantum systems is an important topic at present as it promises to open up investigations into many new areas of science~\cite{Feynman82,Lloyd96,Georgescu14}. This includes exploring exotic states of matter~\cite{Wen07}, thermalisation and equilibration processes~\cite{Eisert15,Nand15}, chemical reaction dynamics~\cite{Kassal11} and probing quantum effects in biological systems~\cite{Lambert13,Dorner12}. A number of approaches are currently being studied, using both classical and quantum methods. While classical methods are limited to specific conditions for efficient simulation of quantum systems~\cite{Foulkes01,Scholl11}, quantum methods have a much larger scope, and a range of techniques have been developed so far, such as analogue~\cite{Feynman82,Lloyd96}, digital~\cite{Montanaro16,Sweke14}, digital-analogue~\cite{Mezz14,Arraz16}, algorithmic~\cite{Kreula16,Kreula16b,Bauer16} and embedded~\cite{Pedernales14,Zhang15}, each with its own advantages and disadvantages. Most methods consider ideal quantum systems, where the constituent elements are isolated from the outside world. However, realistic quantum systems invariably interact with some environment~\cite{Breuer02}. Work on modelling and simulating such quantum systems has seen much progress recently~\cite{Blatt12,Aspuru12,Muller12}, and may shed light on fundamental physical phenomena, including phase transitions in dissipative systems~\cite{Baumann10,Diehl10,Tomadin11}, thermalisation~\cite{Znid10,Znid15} and using dissipation as a resource~\cite{Kraus08,Vert09}. In this context, the development of techniques to realise quantum channels~\cite{Caruso14,Iten16} representing the dynamics of realistic quantum systems has seen rapid growth -- most notably for single qubits~\cite{Wang13,Sweke14b,Bongioanni10, Fisher12, Liu11, Chiuri12, Lu15, Cialdi17} and qudits~\cite{Wang15,Piani11, Marques15}. So far, however, studies have been limited to the standard quantum circuit model~\cite{Nielsen00}.

A natural model for simulating quantum systems is the measurement-based model~\cite{Raussendorf01,Raussendorf03,Briegel09}, which has been used to demonstrate the simulation of quantum computing on entangled resource states using only single-qubit measurements~\cite{Walther05, Prevedel07,Tame07,Vallone10,Lee12,Tame14,Barz11}. The measurement-based model is an interesting method for simulating quantum systems, as it can do this simply by carrying out quantum computing~\cite{Brown10}. However, there may also be the possibility of going further by exploiting the structure of the entangled resource being used to reduce the overall complexity and put a given simulation within reach of current technology. Recently, the first steps in this direction have been taken theoretically~\cite{Bermejo17}. Despite this potential, the realisation and simulation of realistic quantum systems using the measurement-based model has not yet been explored.

In our work we address this issue by introducing and experimentally demonstrating a method for the realisation of a quantum channel that can be used to represent the dynamics of a realistic quantum system using the measurement-based model. We demonstrate the simple case of a single qubit. To do this, we find an efficient mapping from the circuit model to the measurement-based model for the simulation, which allows us to consider the use of an entangled linear cluster state of only four qubits made from three photons -- using the polarisation degree of freedom of each photon as a qubit and the path degree of freedom of one of the photons as an additional qubit. Many previous photonic experiments using cluster states have employed only the polarisation degree of freedom to carry out quantum protocols~\cite{Walther05, Prevedel07,Tame07,Vallone10,Lee12,Barz11}, however the use of other degrees of freedom to represent qubits in `hybrid' cluster states has been considered in order to improve the state quality and protocol results~\cite{Chen07,Vallone07,Park07,Vallone08,Kal10}. In recent work, a quantum error-correction code~\cite{Bell14a}, a secret-sharing protocol~\cite{Bell14b} and a quantum algorithm~\cite{Tame14} have all been realised using four-photon cluster states consisting of both polarisation and path qubits. The setup we use is similar to these experiments, however the overall goal is different and the use of only three photons compared to four ensures we can achieve a high quality performance for our measurement-based realisation of a quantum channel.

By measuring the qubits of our hybrid cluster state in a particular way we are able to realise arbitrary damping channels on a logical qubit residing within the cluster state. The main advantage of this measurement-based approach over the standard circuit model~\cite{Bongioanni10, Fisher12, Liu11, Chiuri12, Lu15, Cialdi17} is that only the pattern of measurements needs to be modified in order to implement different system dynamics. This is particularly useful in a photonic setting, where a reconfiguring of the basic optical elements is not required, both in bulk~\cite{Walther05, Prevedel07,Tame07,Vallone10,Lee12,Tame14,Barz11} and on-chip setups~\cite{Silverstone15,Ciampini16,Schwartz16}. The experimental results obtained match the theoretical expectations well and highlight the potential use of the measurement-based model as an alternative approach to realising quantum channels.
\begin{figure*}[t]
\centering
\includegraphics[width=17cm]{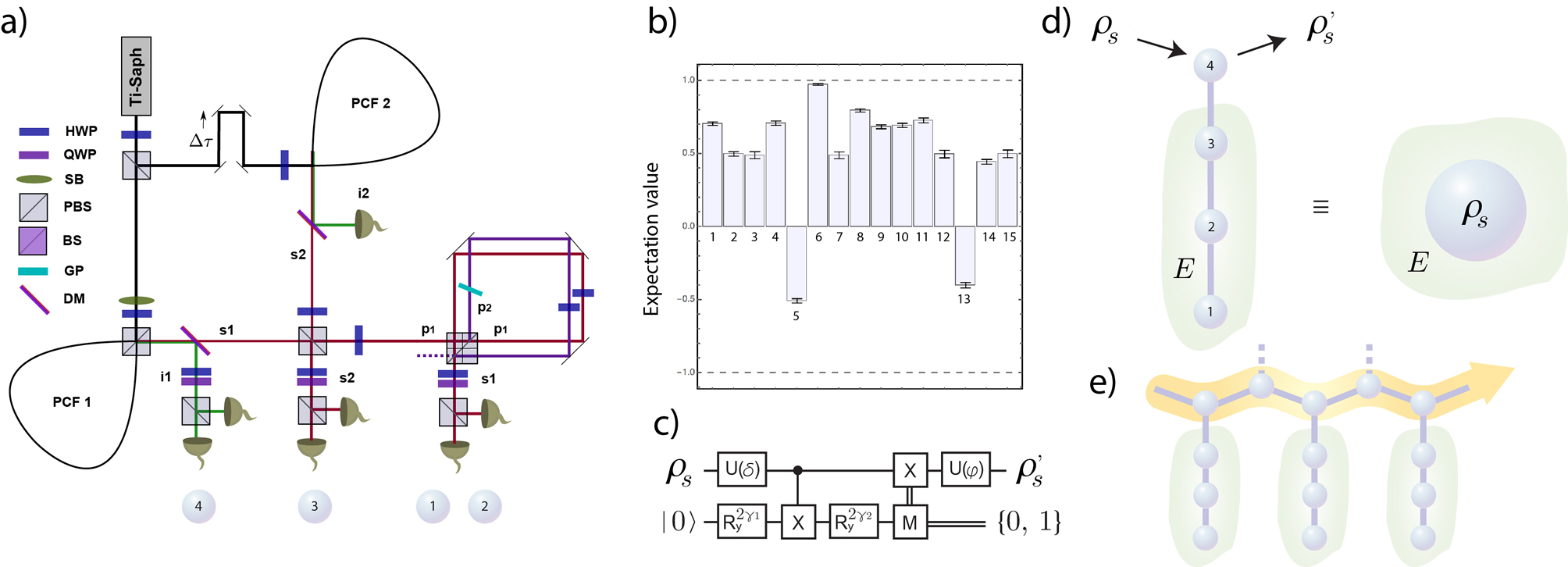}
\caption{Experimental scheme for realising a quantum channel for a single qubit using the measurement-based model. {\bf (a)} Experimental photonic setup, with photonic crystal fibers (PCFs), half-wave plates (HWPs), quarter-wave plates (QWPs), Soleil-Babinet (SB), polarizing beamsplitter (PBS), beamsplitter (BS), glass plate (GP) and dichroic mirror (DM). The setup generates a four-qubit cluster state between photons $s1$, $s2$ and $i1$, with the polarisation and path degree of freedom of photon $s1$ used to represent two qubits. {\bf (b)} Expectation values used for calculating the quality of the generated cluster state in terms of the fidelity. {\bf (c)} Circuit model for simulating an arbitrary single-qubit channel. {\bf (d)} Measurement-based protocol for implementing the simulation of the channel and its equivalent representation. {\bf (e)} Scheme for generalizing the approach to a full open quantum system simulation for a single qubit, where rotations and/or interactions with other qubits (dotted lines) can be carried out stroboscopically.}
\label{fig1} 
\end{figure*}

%%%%%%%%%%%%%%%%%%%%%%%%%%%%
%%%%%%%%%%%%%%%%%%%%%%%%%%%%
%%%%%%%%%%%%%%%%%%%%%%%%%%%%
%%%%%%%%%%%%%%%%%%%%%%%%%%%%

{\it Experimental setup.---} The experimental setup is shown in Fig.~\ref{fig1}~(a). It generates a four-qubit linear cluster state made of three photons -- three qubits are encoded in the polarisation degree of freedom of three photons using the basis $\{\ket{H},\ket{V}\}$, and the fourth qubit is encoded in the path degree of freedom of one of the photons using the basis $\{\ket{p_1},\ket{p_2} \}$. The photons are generated by spontaneous four-wave mixing in photonic crystal fibers (PCFs) tailored to generate a spectrally separable naturally narrowband bi-photon state cross-polarised to the pump~\cite{Fulconis07,Halder09}. The signal wavelength is $\lambda_s \approx 625$~nm and the idler wavelength is $\lambda_i \approx 876$~nm when the PCF is pumped at $\lambda_p=726$~nm. For the pump laser a 80~MHz repetition rate femto-second Ti-Sapphire laser is filtered through a 4F arrangement, with a spectral mask on the Fourier plane achieving the desired spectrum with bandwidth $\Delta \lambda_p = 1.7$~nm to minimise parasitic non-linear effects which reduce photon purities~\cite{Bell15}, and sent to two PCF sources. One of the PCF sources (PCF 1) is arranged in a twisted Sagnac-loop configuration to generate the polarisation entangled Bell pair $\frac{1}{\sqrt{2}}(\ket{HH}+\ket{VV})_{s1\,i1}$ on the signal and idler photons $s1$ and $i1$, respectively, for which we achieve Bell state fidelities of 0.89, limited by the spectral separability of the generated photon pairs~\cite{Fulconis07}. The second source (PCF 2) is pumped in one direction only with the generated state state $\ket{H}_{s2}\ket{H}_{i2}$~\cite{Halder09}. The idler photon, $i2$, serves as a heralding photon for the successful generation of the signal photon, $s2$. After each PCF, the signal and idler photons are separated by dichroic mirrors (DM) and bandpass filtered with widths $40$~nm and $10$~nm respectively to remove Raman noise. The signal photon $s2$ is rotated by a half-wave plate (HWP) into the state $\ket{+}=\frac{1}{\sqrt{2}}(\ket{H}+\ket{V})/\sqrt{2}$ and overlapped with the signal photon $s1$ at a polarising beam splitter (PBS), with the relative arrival time set by the pump delay so that $\Delta \tau \rightarrow 0$. When one signal photon exits each port of the PBS the heralded state is the three-qubit GHZ state in the polarisation bases of the photons $s1$, $s2$, and $i1$: $\frac{1}{\sqrt{2}}(\ket{HHH}+\ket{VVV})_{s1\,s2\,i1}$. The quality of this `fusion' operation is however, limited by the spectral-temporal indistinguishability of the signal photons generated in each source, which can be mitigated to some extent by temperature tuning one of the sources, but limits the fidelity of the three-qubit GHZ state to $0.80\pm0.01$~\cite{McCutcheon16}

The three photons are collected into single-mode fibers, from which $s2$ and $i1$ are sent straight to tomography stages consisting of automated quarter-wave plates (QWPs) and HWPs, followed by PBSs and pairs of single-photon avalanche photodiode detectors (APDs) capable of performing projective measurements onto arbitrary polarisation bases~\cite{James01}. The signal photon $s1$ is path expanded to encode the fourth qubit. This entails a folded Mach-Zehnder interferometer (FMZI) with the anticlockwise and clockwise paths corresponding to the eigenstates of the path qubit, $\ket{p_1}$ and $\ket{p_2}$~\cite{Tame14}. When the incoming photon meets the PBS on entering the FMZI, it performs a controlled-not operation between the polarisation qubit and the path qubit of photon $s1$. With the addition of a HWP before and after the PBS to perform Hadamard operations on the polarisation, the state generated is equivalent to a four-qubit linear cluster state 
\be
\ket{\psi}=\frac{1}{2}(\ket{+00+}+\ket{+01-}+\ket{-10+}-\ket{-11-})_{1\,2\,3\,4},
\ee
where we have written all qubits in the computational basis and a Hadamard operation has also been applied to the polarisation of photon $i1$, performed at the tomography stage. Here, qubit 1 is represented by the polarisation of photon $s1$, qubit 2 by the path of photon $s1$, qubit 3 by the polarisation of photon $s2$ and qubit 4 by the polarisation of photon $i1$. To achieve arbitrary projective measurements for the polarisation qubit of photon $s1$, we use a tomography stage as described for photons $i1$ and $s2$. For the path qubit of photon $s1$, to achieve computational basis measurements we alternate blocking of the paths in the FMZI so that the population of photons in paths $\ket{p_1}$ or $\ket{p_2}$ can be measured after the paths are merged on a 50:50 beamsplitter (BS). Basis measurements on the equatorial plane of the Bloch sphere are achieved by imparting a relative phase between the paths in the FMZI using a glass plate (GP) mounted on an automated rotation stage, followed by the Hadamard operation achieved by the paths combining on the BS~\cite{Tame14}. By using a dual PBS-BS cube for the FMZI the relative path length and therefore phase between the paths can be made relatively stable, leading to an interference visibility of $0.93$ with heralded single photons.

The cluster state $\ket{\psi}$ is the state generated in our setup in the ideal case. However, due to the various dominant sources of error discussed above, including spectral and spatial imperfections introduced during the four-wave mixing process at the PCFs~\cite{Fulconis07,Halder09,McCutcheon16,Bell15}, the fusion PBS between the signal photons~\cite{Bell12}, the path expansion~\cite{Tame14} and to a lesser extent higher-order photon emissions and fibre inhomogeneity~\cite{McCutcheon16,Bell12}, the actual state generated is a mixed state. We therefore first characterise the quality of the cluster state generated in our setup. The fidelity $F={\rm Tr}(\rho_{exp}\ket{\psi}\bra{\psi})$ quantifying the overlap between the experimental state $\rho_{exp}$ and the ideal state $\ket{\psi}$ can be obtained by decomposing the projector $\ket{\psi}\bra{\psi}$ into a summation of terms made from projector elements arising from the eigenvectors of tensor products of Pauli operators. Each term can then be measured locally, with the total expectation value of all the terms for $\rho_{exp}$ giving the fidelity. There are a total of 15 terms~\cite{Toth05}, leading to a fidelity of $F=0.63 \pm 0.01$. The expectation values of the terms are shown in Fig.~\ref{fig1}~(b). The presence of genuine multipartite entanglement, signifying that all qubits were involved in the generation of the state, is confirmed as $F>0.5$~\cite{Toth05}. Improvements to the quality of our state could be made by operating at a reduced pump power for the PCFs in order to suppress higher-order photon emissions from the four-wave mixing~\cite{McCutcheon16,Bell12}. However, this reduces the state generation rate and impacts on the data collection time. Better matching of the spectral profiles of the signal photons produced via four-wave mixing processes in separate PCFs would also improve the state quality as the PBS fusion operation relies on spectral indistinguishability of the photons~\cite{Bell12}. While the above factors would improve the state quality, the current fidelity value is comparable to other photonic cluster state experiments and allows us to demonstrate a proof-of-principle realisation of a quantum channel using the measurement-based model.

%%%%%%%%%%%%%%%%%%%%%%%%%%%%
%%%%%%%%%%%%%%%%%%%%%%%%%%%%
%%%%%%%%%%%%%%%%%%%%%%%%%%%%
%%%%%%%%%%%%%%%%%%%%%%%%%%%%

{\it Results.---} We start our implementation by showing how the standard circuit model for realising a quantum channel is mapped to the measurement-based model. In Fig.~\ref{fig1}~(c) the quantum circuit for carrying out an arbitrary completely positive trace-preserving (CPTP) channel for a single qubit $\rho_S$ is depicted~\cite{Wang13}. For simplicity, the unitary operations $U(\delta)$ and $U(\varphi)$ at the start and end are not considered, as they are not needed for the specific examples we demonstrate. They are local operations and if needed for a given channel they can be applied easily in the measurement-based model~\cite{Briegel09}. In the circuit, the rotation $R_y^{\theta}=\left(\begin{array}{cc} \cos \theta/2 & -\sin \theta/2 \\ \sin \theta/2 & \cos \theta/2 \end{array}\right)$, $X$ is the Pauli $\sigma_x$ operator and $M$ represents a measurement in the computational basis. The circuit shown implements the quantum channel ${\cal E} (\rho_S) \to K_0\rho_S K_0^\dag+K_1\rho_S K_1^\dag$, where the Kraus operators are $K_0=\left(\begin{array}{cc} \cos \beta & 0 \\ 0 & \cos \alpha \end{array}\right)$ and $K_1=\left(\begin{array}{cc} 0 & \sin \alpha \\ \sin \beta & 0 \end{array}\right)$. The relations linking these operators to the rotations in the circuit are $\gamma_1=(\beta-\alpha+\pi/2)/2$ and $\gamma_2=(\beta+\alpha-\pi/2)/2$. If the measurement outcome $M$ of the ancilla qubit is 0, then the operator $K_0$ is applied and if it is 1, then an $X$ operation is applied to the system qubit in order for the operator $K_1$ to be applied~\cite{Wang13}. Taking into account that both outcomes can occur for the ancilla qubit measurement, the system is put into a summation of the two processes. We stress that this procedure is capable of simulating arbitrary single-qubit channels of which there exist a continuous family. In this work we will demonstrate 3 different channels: amplitude damping, phase damping and a channel we call $\beta$ damping, an example extremal channel characterised by simultaneous amplitude damping and phase damping occurring in perpendicular bases. For the first two channels it is convenient to set the parameters in the circuit as $\alpha=\cos^{-1}(e^{-\eta t/2})$ and $\beta=0$, where $\eta$ is an effective damping rate and $t$ is the simulation time desired. Amplitude damping is then implemented naturally by the circuit. On the other hand, phase damping does not require the $X$ operation from the ancilla measurement outcome 1 to be applied. For the third channel we fix $\alpha$ and choose specific values of $\beta$. We now map the circuit model to the measurement-based model and show that a four-qubit entangled cluster state is all that is needed to carry out the simulation. While we do not claim that our mapping is optimal, in that it may be possible to do some elements of the simulation using only a three-qubit cluster state, the efficient mapping we present puts the simulation within reach of our setup and allows us to experimentally demonstrate the fundamental workings of a measurement-based approach.

The measurement-based model involves making single-qubit measurements on a cluster state in order to carry out logic operations on quantum information encoded within. For cluster states two types of measurements allow logic operations to be performed: (i) Measuring a qubit $j$ in the computational basis allows it to be disentangled and removed from the cluster, leaving a smaller cluster of the remaining qubits, and (ii) In order to perform logic gates, qubits must be measured in the equatorial basis $B_j(\alpha)=\{ \ket{\alpha_+}_j,\ket{\alpha_-}_j \}$, where $\ket{\alpha_{\pm}}_j=(\ket{0}\pm e^{-i \alpha}\ket{1})_j/\sqrt{2}$, for $\alpha\!\in\!(0,2\pi]$. This measurement on qubit $j$, initially in the logical state $\ket{\phi}$, results in propagation of the state to qubit $j+1$ with the operations $\sigma_x^s {\sf H} R_z^\alpha$ applied. Here, the rotation $R_z^\alpha={\rm exp}(-i \alpha \sigma_z/2)$ has been applied along with a Hadamard operation, ${\sf H}$, and a Pauli $X$ operation dependent on the outcome $s$ from the measurement~\cite{clusterback}.
\begin{figure*}[t]
\centering
\includegraphics[width=17.5cm]{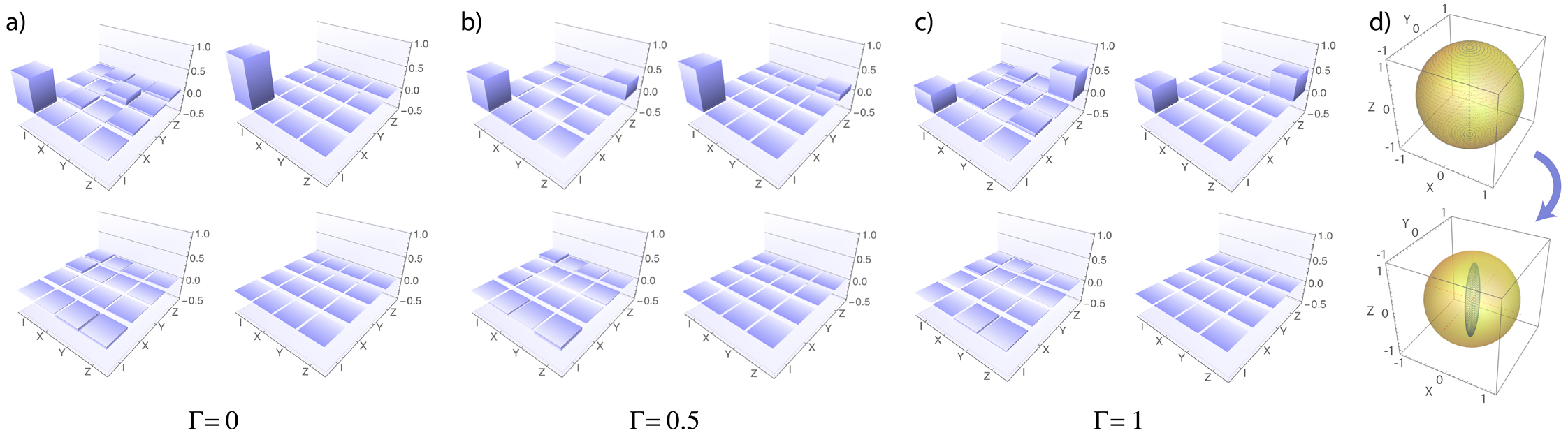}
\caption{Quantum process matrices for the realisation of a phase damping channel. {\bf (a)} $\Gamma=0$. {\bf (b)} $\Gamma=0.5$. {\bf (c)} $\Gamma=1$. {\bf (d)} Bloch sphere representation showing the effect of the channel at $\Gamma=1$. In panels (a)-(c) the left column is the experimental result and the right column is the ideal case, with the top row corresponding to the real part and the bottom row to the imaginary part of the elements of the matrix.}
\label{fig2} 
\end{figure*}

\begin{figure}[b]
\centering
\includegraphics[width=9cm]{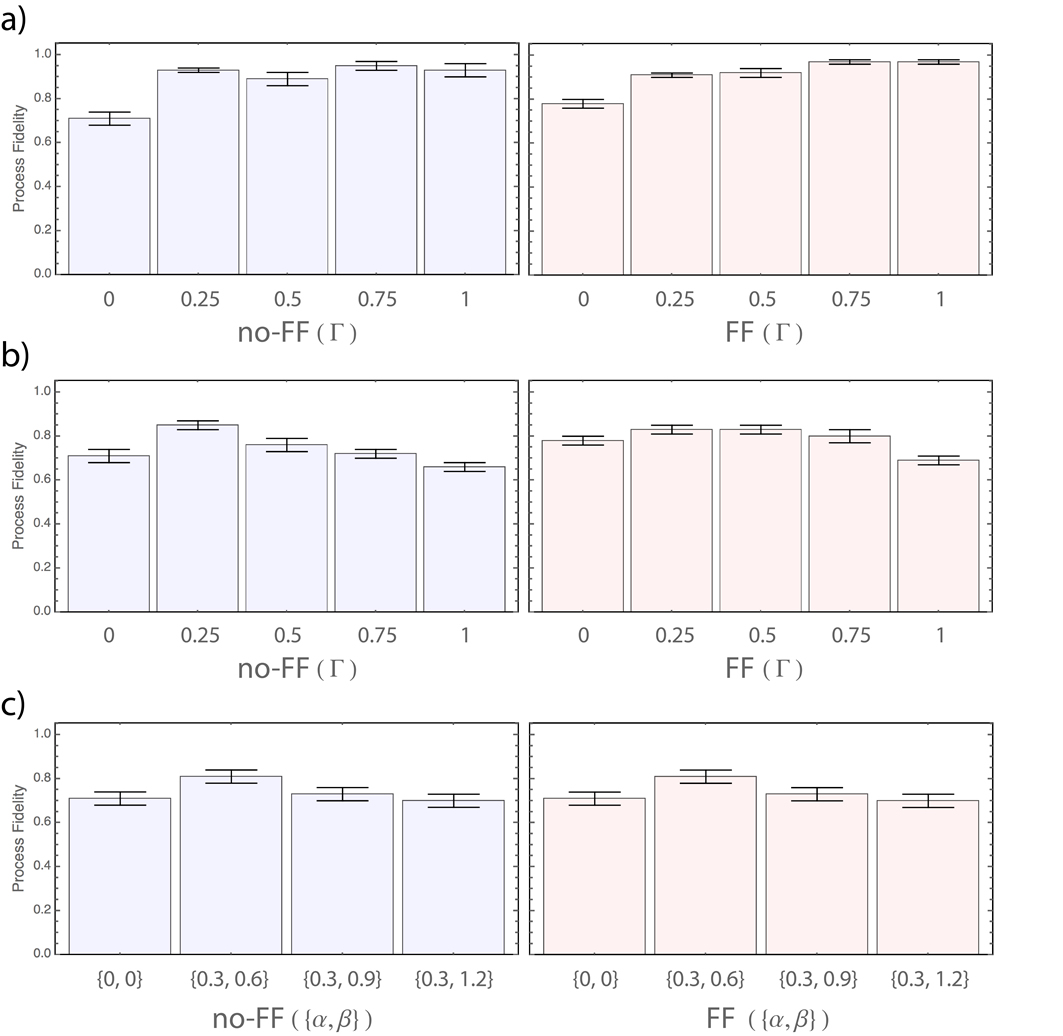}
\caption{Process fidelities for the realisation of phase, amplitude and $\beta$ damping channels. {\bf (a)} Phase damping. {\bf (b)} Amplitude damping. {\bf (c)} $\beta$ damping. In all, the first column corresponds to the case $s_1=0$ and $s_2=0$ for the outcomes of the measurements of the qubits in the cluster, no feed-forward (no-FF), while the second column corresponds to the case $s_1=0$ and $s_2=1$, feed forward (FF), with appropriate byproduct operator applied to the output.}
\label{fig3} 
\end{figure}

\begin{figure*}[t]
\centering
\includegraphics[width=17.5cm]{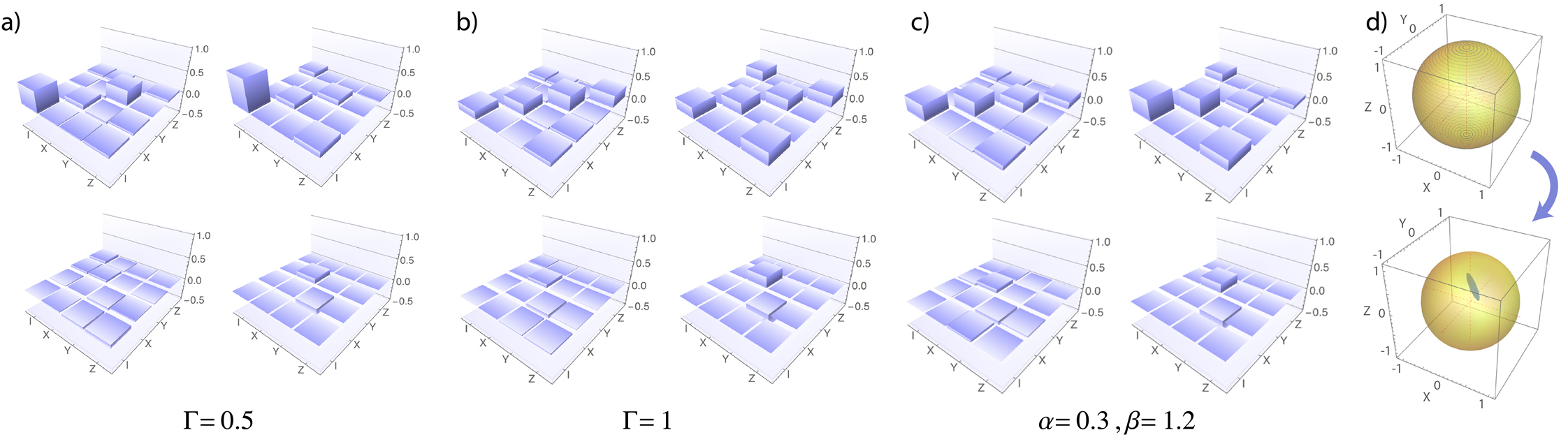}
\caption{Quantum process matrices for the realisation of an amplitude damping and $\beta$ channel. {\bf (a)} $\Gamma=0.5$ for amplitude damping. {\bf (b)} $\Gamma=1$ for amplitude damping. {\bf (c)} $\alpha=0.3$ and $\beta=1.2$ for rotated phase damping. {\bf (d)} Bloch sphere representation showing the effect of the amplitude damping channel at $\Gamma=1$. In panels (a)-(c) the left column is the experimental result and the right column is the ideal case, with the top row corresponding to the real part and the bottom row to the imaginary part of the elements of the matrix.}
\label{fig4} 
\end{figure*}

Using the cluster state generated in our experiment, the input states corresponding to the ancilla qubit $\ket{+}$ and system qubit $\rho_S=\ket{+}\bra{+}$ are naturally encoded on qubits 1 and 4, respectively, as shown in Fig.~\ref{fig1}~(d). For the ancilla qubit, note that in the circuit model shown in Fig.~\ref{fig1}~(c), the first gate, $R_y^{2\gamma_1}$, is applied to an initial state $\ket{0}$. This gate can be decomposed into a product of several gates: $R_z^{\pi/2}{\sf H} R_z^{2\gamma_1} {\sf H} R_z^{-\pi/2}$. Taking the first two operations of the gate, we have ${\sf H} R_z^{-\pi/2} \ket{0}=\ket{+}$. Therefore the remaining operations that need to be carried out on the ancilla qubit using the cluster state are $R_z^{\pi/2}{\sf H} R_z^{2\gamma_1}$. By including the controlled-{\sf X} (${\sf CX}$) gate between the ancilla and system, and the subsequent gate $R_y^{2\gamma_2}$, the total operation for the remainder of the circuit for both system and ancilla is given by $(\openone \otimes R_y^{2\gamma_2}{\sf H}) {\sf CZ} (\openone \otimes {\sf H} R_z^{\pi/2}{\sf H} R_z^{2\gamma_1})$, where the system is the first qubit and the ancilla is the second. Here, we have decomposed the ${\sf CX}$ gate as $(\openone \otimes {\sf H}){\sf CZ}(\openone \otimes {\sf H})$. The first two operations, ${\sf H} R_z^{2\gamma_1}$, are implemented by measuring qubit 1 of the cluster state in the basis $B_1(2\gamma_1)$, which propagates the logical ancilla to qubit 2 of the cluster. The next two operations, ${\sf H} R_z^{\pi/2}$, are implemented by measuring qubit 2 of the cluster state in the basis $B_2(\pi/2)$, which propagates the ancilla to qubit 3. The {\sf CZ} gate is then naturally applied as the logical qubit of the ancilla now resides on qubit 3 and the logical qubit of the system resides on qubit 4 -- the edge linking qubits 3 and 4 is a {\sf CZ} gate. The final two operations, $R_y^{2\gamma_2}{\sf H}$, are incorporated into the measurement basis of the ancilla qubit on qubit 3, which is normally measured in the computational basis. Thus, we have outcomes $\{0,1\}$ of the ancilla in the circuit corresponding to the outcomes of the measurement basis $\{ {\sf H} (R_y^{2\gamma_2})^\dag \ket{0},{\sf H} (R_y^{2\gamma_2})^\dag \ket{1} \}$. 

In the measurement-based model it is important to include the unwanted Pauli byproduct operators that act on the logical qubits due to the random nature of the outcomes of measurements of qubits in the cluster. Including the byproducts makes the logical operations fully deterministic~\cite{Briegel09}. The byproducts can be propagated right to the end and incorporated into the final measurements of the system and ancilla. For the ancilla, the byproducts lead to the measurement basis $\{ \sigma_z^{s_1}\sigma_x^{s_2} {\sf H} (R_y^{2\gamma_2})^\dag \ket{0}, \sigma_z^{s_1}\sigma_x^{s_2}{\sf H} (R_y^{2\gamma_2})^\dag \ket{1} \}$ for qubit 3, and the basis of qubit 2 must be modified to $B_2((-1)^{s_1}\pi/2)$. Here, $s_i$ corresponds to the measurement outcome for qubit $i$. For the system, the byproduct operation is $\sigma_z^{s_2}\sigma_x^{s_3}$, where the $\sigma_x$ from measurement $M$ in the circuit has been included. All the operations from the circuit model have now been mapped into the measurement-based model and it is clear that a four-qubit linear cluster state is sufficient for simulating the action of an arbitrary channel on a single qubit. 

To generalise this method to a full simulation of a single-qubit quantum system, one might also like to include interaction with additional systems or a rotation while it is being subject to damping. In this case, the interaction/rotation and damping could be split up into smaller time steps and carried out stroboscopically as the logical qubit propagates along a larger cluster state (taking into consideration the passage of byproducts through the corresponding circuit), as highlighted in Fig.~\ref{fig1}~(e). Furthermore, the simulation of channels with memory effects could be included by conditioning future time steps on the outcome of the ancilla measurement, $s_3$, or initially entangling the ancilla qubits in order to introduce correlations in the environmental degrees of freedom~\cite{Caruso14}.

We now characterise the performance of the measurement-based approach for phase damping using the cluster state generated in our setup. We choose the basis states of our qubit to simulate that of a two-level system: $\{ \ket{g},\ket{e}\}$. For this, we use the convention $\ket{H}=\ket{0}\leftrightarrow \ket{e}$ and $\ket{V}=\ket{1}\leftrightarrow \ket{g}$, and combine the damping rate and time into a single quantity, $\Gamma$, with the correspondence $ \sqrt{1-\Gamma}=e^{\eta t/2}$. We then choose 5 different damping values: $\Gamma=\{0, 0.25, 0.5, 0.75, 1 \}$. These values determine the parameters $\alpha=\cos^{-1}(\sqrt{1-\Gamma})$ and $\beta=0$, which are inserted into the formulas for $\gamma_1$ and $\gamma_2$ to obtain the angles for the measurements of qubits in the cluster. For each value of $\Gamma$, we carry out quantum process tomography~\cite{Chuang97} by encoding the probe states $\ket{g}$, $\ket{e}$, $\ket{+}$ and $\ket{+_y}=\frac{1}{\sqrt{2}}(\ket{g}+i\ket{e})$, and perform quantum state tomography on the output of the channel for each probe state~\cite{James01}. From this information we reconstruct the process matrix $\chi$ for the channel, defined by the relation ${\cal E}(\rho_S)=\sum_{i,j} \chi_{ij} E_i \rho_S E_j^\dag$, with the operators $E_i$ forming a complete basis for the Hilbert space, $E_i=\{\openone,X, Y, Z \}$~\cite{Nielsen00}. The probe state $\ket{+}$ is naturally encoded into the cluster state, whereas the probe state $\ket{+_y}$ is encoded using a QWP on photon $i1$, and the probe states $\ket{g}$ and $\ket{e}$ are encoded using a polariser. In Fig.~\ref{fig2}~(a), (b) and (c) we show the $\chi$ matrices for the simulation of phase damping for $\Gamma=0$, $0.5$ and $1$, respectively, for the case of measurement outcomes $s_1=0$ and $s_2=0$. The left column in each corresponds to the experiment, $\chi$, and the right column the theoretically expected ideal case, $\chi_{id}$. One can see that the process matrices match well, with process fidelities defined as $F_p={\rm Tr}(\sqrt{\sqrt{\chi}\chi_{id}\sqrt{\chi}})^2/{\rm Tr(\chi){\rm Tr}(\chi_{id})}$~\cite{Jozsa94} equal to $0.71\pm 0.03$, $0.89\pm 0.03$ and $0.93\pm 0.03$, respectively. In Fig.~\ref{fig3}~(a) we show $F_p$ for all values of $\Gamma$ simulated. The left hand side (blue columns) shows the case of $s_1=0$ and $s_2=0$, which we call `no feed forward' (no-FF), while the right hand side (red columns) shows the case of $s_1=0$ and $s_2=1$ (FF), chosen as an example of when byproducts are produced and the necessary rotations are applied to $\rho_S$, which are incorporated into the measurements during the state tomography. 

It can be seen in Fig.~\ref{fig3} that there is little difference in the process fidelities of the no-FF and FF cases, which indicates that there is not much bias in the implementation of the channel due to the measurement outcomes of qubits in the cluster state. As FF operations are needed to make the channels fully deterministic in the measurement-based model, the results show that the channels can be carried out deterministically and with consistent performance. While the main quantifier of how well the channels perform can be taken to be the process fidelities shown in Fig.~\ref{fig3}, the $\chi$ matrices shown in Fig.~\ref{fig2} help visualise what the channels are doing in the Pauli operator basis. As an additional complementary plot, in Fig.~\ref{fig2}~(d) we show the effect of the channel on the Bloch sphere for $\Gamma=1$. The Bloch sphere is squashed into a cigar shape along the $z$-axis as expected~\cite{Nielsen00}.

In Fig.~\ref{fig4}~(a) and (b) we show the $\chi$ matrices for the simulation of amplitude damping for $\Gamma=0.5$ and $\Gamma=1$. The $\chi$ matrix for $\Gamma=0$ is the same as the phase damping channel. The process fidelities for these channels are $0.76 \pm 0.03$ and $0.66 \pm 0.02$. The full range of process fidelities is given in Fig.~\ref{fig3}~(b) for the no-FF and FF cases. In Fig.~\ref{fig4}~(d) we show the effect of the channel on the Bloch sphere for $\Gamma=1$. The Bloch sphere is squashed into a cigar shape, similar to the phase damping case, but at the same time it is gradually pushed toward the basis state $\ket{g}$, as expected~\cite{Nielsen00}. In Fig.~\ref{fig4}~(c) we show an example $\chi$ matrix for the case when $\beta \neq 0$, which we call the `$\beta$ channel'. Here, we have set $\alpha=0.3$ and $\beta=1.2$. The corresponding process fidelity is $0.70 \pm 0.03$. In Fig~\ref{fig3}~(c) we show the process fidelities for other non-zero $\beta$ values, both in the no-FF (left hand side) and FF cases (right hand side). The $\beta$ channel results show that the measurement-based method can be used for simulating non-standard quantum channels representing realistic quantum system dynamics.

%%%%%%%%%%%%%%%%%%%%%%%%%%%%
%%%%%%%%%%%%%%%%%%%%%%%%%%%%
%%%%%%%%%%%%%%%%%%%%%%%%%%%%
%%%%%%%%%%%%%%%%%%%%%%%%%%%%

{\it Discussion.---} In this work we experimentally demonstrated a method for the realisation of quantum channels using the measurement-based model for the simple case of a single qubit. We mapped the circuit model to the measurement-based model and showed that an entangled linear cluster state of only four qubits made from three photons is sufficient. By measuring the qubits of the cluster state we were able to simulate different quantum channels, including amplitude and phase damping, on a logical qubit residing within the cluster state. The experimental results match the theoretical expectations well. We also briefly discussed how to extend the method to implement a full simulation of a single-qubit quantum system that would include rotations while the decoherence takes place. Our results highlight the potential use of the measurement-based model as an alternative approach to simulating realistic quantum systems. Future work could look into whether a smaller cluster state of only two or three qubits can also be used for demonstrating specific quantum channels. In addition, it would be interesting to see how extra qubits provide extended functionality and flexibility. Furthermore, one could extend the model to qudits, collective multiqubit channels and even memory effects.

%%%%%%%%%%%%%%%%%%%%%%%%%%%%
%%%%%%%%%%%%%%%%%%%%%%%%%%%%
%%%%%%%%%%%%%%%%%%%%%%%%%%%%
%%%%%%%%%%%%%%%%%%%%%%%%%%%%

{\it Acknowledgments.---} This work was supported by the UK's Engineering and Physical Sciences Research Council (EP/L024020/1), EU FP7 grant 600838 QWAD, the South African National Research Foundation and the South African National Institute for Theoretical Physics.

%%%%%%%%%%%%%%%%%%%%%%%%%%%%
%%%%%%%%%%%%%%%%%%%%%%%%%%%%
%%%%%%%%%%%%%%%%%%%%%%%%%%%%
%%%%%%%%%%%%%%%%%%%%%%%%%%%%

\end{document}